# The impact factor's Matthew effect: a natural experiment in bibliometrics

Vincent Larivière and Yves Gingras

Observatoire des sciences et des technologies (OST)
Centre interuniversitaire de recherche sur la science et la technologie (CIRST)
Université du Québec à Montréal (UQAM)
CP 8888, Succursale Centre-ville
Montréal, Québec, H3C 3P8
Email: [lariviere.vincent; gingras.yves]@uqam.ca

**Abstract**
Since the publication of Robert K. Merton's theory of cumulative advantage in science (Matthew Effect), several empirical studies have tried to measure its presence at the level of papers, individual researchers, institutions or countries. However, these studies seldom control for the intrinsic "quality" of papers or of researchers—"better" (however defined) papers or researchers could receive higher citation rates because they are indeed of better quality. Using an original method for controlling the intrinsic value of papers—identical duplicate papers published in different journals with different impact factors—this paper shows that the journal in which papers are published have a strong influence on their citation rates, as duplicate papers published in high impact journals obtain, on average, twice as much citations as their identical counterparts published in journals with lower impact factors. The intrinsic value of a paper is thus not the only reason a given paper gets cited or not; there is a specific Matthew effect attached to journals and this gives to paper published there an added value over and above their intrinsic quality.

**Introduction**
In a paper published in *Science* about forty years ago (Merton, 1968), the sociologist Robert K. Merton suggested that recognition for a discovery is more easily attributed to scientists already recognized than to lesser known scientists. Quoting the Gospel according to St. Matthew "For to all those who have, more will be given, and they will have an abundance; but from those who have nothing, even what they have will be taken away" (Matthew 25/29), he named that phenomena "Matthew effect". Since the publication of that seminal paper, empirical evidence of the Matthew effect has been found at several levels of aggregation of the science system (Katz, 1999), from individual researchers (Costas, Bordons, van Leeuwen and van Raan, 2009; Tol, 2009) to institutions (Medoff, 2006) and countries (Bonitz, Bruckner & Scharnhorst, 1997). More specifically, Tol (2009) provided empirical evidence of a Matthew effect for a sample of prolific economists, using a method based on the growth of firms and their deviation from the Pareto distribution. He shows that highly-cited papers and researchers have increasing returns to scale, as their citation rates grow faster than the Pareto distribution. In the same field, Medoff (2006) found that papers authored by researchers from elite institutions (Harvard University and the University of Chicago) obtained higher citation rates, even after controlling for author and journal quality and characteristics of the article (collaboration, number of pages, etc.). Laudel (2006), studying German and Australian physicists, also provided evidence of a Matthew Effect in research funding, where researchers already funded were more likely to be rewarded with more funding. Another manifestation of the Matthew effect is the non-linear relationship between citations and publications, where $X$ number of publications leads to $X^n$ citations (Katz, 1999; van Raan, 2008a,b). As noted by Bonitz, Bruckner & Scharnhorst (1997): "A minority of countries, expecting a high number of citations per scientific paper, gains more citations than expected, while the majority of countries, expecting only a low number of citations per scientific paper, achieves less citations than expected" (p.408). This measure, however, does not correct for the "quality bias"; countries that gain more citations than expected might in fact be doing better research than those who are losing citations.

Most attempts to measure the "quality" of papers or the recognition of scientists use either citations received or the impact factor of the journal as indicators. The problem however is that these two measures may be correlated and thus it is difficult to distinguish the intrinsic quality of a paper independently of he effect of the journal in which it is published. It is usually believed that it is the citations received by the papers that contribute to the impact factor of the journal (Garfield, 1973). This is certainly true by definition as the impact factor of a journal is calculated using the citations received by the papers contained in that journal. From a sociological point of view, however, one cannot exclude the possibility that citations to a paper are in fact influenced by the impact factor of the journal, a number often considered as a good indicator of the quality and visibility of the journal. It is thus possible that the citations received by a given paper do not only reflect the 'quality' of that paper but also that of the journal in which it is published. This interaction between journal and paper is difficult to measure since we usually cannot keep constant one of the two variables as papers are supposed to be unique contributions and no two papers are supposed to be identical.

This paper solves this problem by taking advantage of the existence of "duplicate" articles published in different journals. The existence of duplicate papers thus offer a "natural experiment" that makes possible to isolate the role of the impact factor of a journal on the citations received by a given paper as duplicates are in fact the very same papers, published in two different journals. By construction, these pairs of papers are of the same "quality" and any significant difference in citations to duplicates must be attributed to the journal itself, whose "quality" or at least "visibility" is measured by the impact factor (IF). The only study of citations to duplicate papers is that of Knothe (2006), who presented the citation measures of four different cases of duplicate or highly related papers in the field of chemistry and biochemistry. Although one could argue that it is not easy to draw a firm conclusion on such a small sample of papers, Knothe concludes that the citations to individual papers are highly related to the impact factor of the journal in which they were published and that "… 'minor' papers published in the 'right' journal might accumulate a stronger citations record than 'major' papers published in the 'wrong' journal" (p. 1,838)

Although the existence of such duplicate papers raises important ethical questions, they offer a unique occasion to test the effect of the impact factor of the journal on the number of citations received by the papers they publish. Duplicate papers being by definition identical, if their number of citations varies strongly according to the journal in which they are published, then the difference in the level of citation can be attributed to the effect of the journal itself. Duplicate papers here play the role that multiple discoveries made by researchers of different "reputation" play in Merton's original analysis. After having explained how we define and identify duplicate papers, we will use the set of duplicate papers published in journals with *different* impact factors to isolate the effect of the impact factor of the journal on the citations received by each of the duplicate papers.

**Methods**
We use Thomson Reuters' Web of Science (WoS)—which includes the Science Citation Index Expanded, the Social Sciences Citation Index, and the Arts and Humanities Citation Index—to locate duplicate papers and compare their scientific impact. Previous studies (see, among others, Errami and Garner, 2008; Smith Blancett, Flanagin and Young, 1995; Larivière and Gingras, 2009; Sorokina *et al.*, 2006) have used different definitions of a duplicate paper, from one extreme (the exact same text is republished) to the other (e.g. only part of the data is reused, same data but different conclusions, etc.). This paper uses the simple method developed by Larivière and Gingras (2009) identifying duplicates papers as those that are published in two different journals and have the following metadata in common: 1) the exact same title, 2) the same first author, 3) the same number of cited references. Using this method, we have identified 4,918 pairs of papers. For the present study, we use only the 4,532 pairs that have appeared in journals with different impact factors. Papers republished more than twice were excluded for reasons of homogeneity. In about 80% of the cases, these duplicates were published the same year or the year after. Although this method includes both false positives and false negatives, being uniformly applied to all papers in the WoS database, our definition makes possible an analysis of the comparative citation impact of such presumably identical papers published in different journals[1].

The impact factor is a measure of the impact of scientific journals devised by Garfield and Sher (1963). First aimed at evaluating scientific journals, the impact factor is now increasingly used to evaluate research and orient publishing strategies of researchers and has, in this respect, become a measure of the quality or reputation of a journal[2]. Each journal has an impact factor (IF), which is calculated annually by Thomson Reuters based on the number of citations received by a journal relative to the number of papers it published. The specific calculation is as follows (for the year 2008): number of citations received in 2008 by papers published in a given journal in 2006 and 2007, divided by the number of papers published in that journal in 2006 and 2007. In our calculation of the impact factors, the asymmetry between the numerator and the denominator was also corrected. Indeed, in the calculation of their impact factors, Thomson Reuters counts citations received by all document types published (articles reviews, editorials, news items, etc.) but only divides these citations by the number of articles and reviews published, which are considered as "citable" items. This has the effect of artificially increasing the impact factor of journals that publishes a lot of "non research" items (like editorials and letters), which get cited but are not included in the denominator.

**Results**

We found 4,532 pairs of identical papers published in two journals with different impact factor, for a total of 9,064 papers. To examine the effect of impact factors on articles' citation rates, each paper of a pair of duplicates was assigned to either the "highest" or the "lowest" impact factor bin. We thus have 4,532 papers in the highest impact factor bin and the same number in the lowest impact factor bin. The average impact factor for the first bin is more than twice as high as that of the lower impact factor of the second bin (1.11 vs. 0.47) and the range is between 0.007 and 23.3 for the highest impact and 0 and 14.1 for the lowest impact.

Table 1 presents impact measures for the higher impact factor and lower impact factor duplicates. It shows that papers published in journals with the higher impact factor were, on average, twice as much cited (11.9) as their duplicates published in a journal with a lower impact factor (6.33). The median number of citations is also higher for papers in high impact journal (2.74 vs. 1.6), while the percentage of uncited papers is lower. All differences in impact measures are significant at p<0.001 .This strongly demonstrates that the journals in which papers are published have an indisputable influence on their future citation rates, as papers published in high impact journals obtain twice as much citations as their identical counterpart published in a lower impact factor journal.

---

[1] For more details on the method used for detecting duplicate papers, see Larivière and Gingras (2009).
[2] For a historical review of impact factors, see Archambault & Larivière (2009).

Table 1. Impact measures of highest impact and lowest impact duplicates

| Indicator | Higest I.F. | Lowest I.F. |
|---|---|---|
| Average impact factor | 1.11 | 0.47 |
| *Average of relative impact factor* | *1.17* | *0.60* |
| Average number of citations | 11.9 | 6.33 |
| *Average of relative citations* | *1.04* | *0.63* |
| Median number of citations | 2.74 | 1.60 |
| Percentage of uncited papers | 30,5% | 41,1% |
| **Number of papers** | **4,532** | **4,532** |

One might argue that the differences in citations received by duplicates are in fact due to the different disciplines in which the two papers are published. For example, if a duplicate paper is published once in a citation-intensive discipline (e.g. biomedical research) and once in a non citation-intensive discipline (e.g. mathematics), it is more likely that that the first will receives more citations than the second. In order to take into account the different citation practices of various disciplines, we also compiled discipline-normalized impact measures (average of relative impact factor and average of relative citations). Following Schubert and Braun (1986) and Moed, De Bruin and van Leeuwen (1995), these indicators of relative citations and impact factors are normalized by the world average citations and impact factors for each specialty of the same publication year. Average of relative impact factor and average of relative citations values above (or below) one means that the average scientific impact of paper/journal is above (or below) the world average of their respective subfield. Field normalized impact measures presented in Table 1 show that the differences in citations received by duplicates are still important (1.04 vs. 0.63) and significant at $p<0.001$. Another way to show that for identical papers, the one published in a higher impact journal attracts more citations than the other is to look at the distribution of the duplicates in terms of the proportion that is more (or less) cited in higher (or lower) impact journals. We find that in 43.6% of the cases (1,977 pairs) the paper published in the journal with the highest impact factor obtained more citations and that in only 22.2% of the cases (1,006 pairs) did the papers with lower impact received more citations; for the other 34.2% of the cases (1,549 pairs) there was no difference in citations, but most of these are in fact uncited papers (914 pairs).

Table 2 presents impact measures (Average number of citations and average of relative citations) of the subset of duplicates published in the same discipline (N=3,573 duplicates, 7,146 papers) for disciplines having at least 30 duplicates pairs (60 papers). This positive relation between citations and impact factors for identical papers is seen in all disciplines of the natural and medical sciences but mathematics, and is highly significant in most disciplines (biomedical research, chemistry, clinical medicine, engineering and technology and physics). It is likely that the difference would also be significant in earth and space sciences category and in biology if the number of papers had been higher. In the social sciences and humanities, the difference is highly significant in the broad discipline of social sciences.

Table 2. Average number of citations and average of relative citations of highest impact and lowest impact duplicates for disciplines with at least 30 duplicates pairs (60 papers)

| Discipline | Average number of citations | | Average of relative citations | | Number of duplicate pairs |
|---|---|---|---|---|---|
| | Higest I.F. | Lowest I.F. | Higest I.F. | Lowest I.F. | |
| Biology | 5.97 | 4.17 | 0.69 | 0.59 | 66 |
| Biomedical Research * | 19.77 | 8.15 | 0.68 | 0.37 | 111 |
| Chemistry* | 5.90 | 5.19 | 0.42 | 0.36 | 272 |
| Clinical Medicine *** | 21.46 | 12.08 | 1.40 | 0.83 | 1,342 |
| Earth and Space | 3.86 | 2.98 | 0.25 | 0.19 | 57 |
| Engineering and Technology*** | 5.02 | 1.78 | 0.97 | 0.41 | 961 |
| Health (Social Sciences) | 5.79 | 5.74 | 1.07 | 0.99 | 34 |
| Mathematics | 5.00 | 5.05 | 0.79 | 0.78 | 74 |
| Physics *** | 10.02 | 3.57 | 0.79 | 0.31 | 571 |
| Psychology | 17.10 | 15.63 | 2.59 | 1.78 | 30 |
| Social Sciences ** | 10.98 | 3.49 | 1.57 | 0.90 | 55 |

*** p<0,001; ** p<0,01; * p<0,05.

**Discussion and Conclusion**

The data presented in this paper show that the impact factor of journals do have a strong effect on the citation rates of otherwise identical papers. Although Garfield (1973) is technically right in stating that it is the citations to individual *articles* that determine the impact factor of the *journal*—and not the opposite—our analysis provide strong evidence of a Matthew effect (Merton, 1968) related to the journal itself and that there is a feedback effect from the journal to the paper. We know that many scientists look up the impact factors of journals in order to choose where to publish; they also tend to read journals with high impact factors in their own field and thus the papers in these journals become more visible. The fact that the value of the impact factor of a journal can be manipulated and influenced by many variables—such as its diffusion and the number of researchers active in the field covered by the journal—does not change the fact that the very location of a paper do influence its reception independently of its intrinsic value. The intrinsic value of a paper is thus not the only reason a given paper gets cited or not. In this sense, there is a specific Matthew effect attached to journals and this gives to paper published there an added value over and above their intrinsic quality. Though such a conclusion had to be expected from a sociological point of view, our data provide the first measure of that effect based on a large scale sample covering all disciplines.

**Acknowledgments**

The authors wish to thank Robert Gagnon, Lorie Kloda, Jean Lebel, Benoit Macaluso, Vladimir Pislyakov and Jean-Pierre Robitaille for their useful comments and suggestions.